\documentclass[a4paper,fleqn,usenatbib]{mnras}

\usepackage{newtxtext,newtxmath}

\usepackage[T1]{fontenc}
\usepackage{ae,aecompl}

\usepackage[pdftex]{graphicx}
\usepackage{amsmath}	
\usepackage{amssymb}	
\usepackage{threeparttable}
\usepackage{comment}

\title[Long-term X-ray variation of the colliding wind Wolf-Rayet binary WR~125]{Long-term X-ray variation of the colliding wind Wolf-Rayet binary WR~125}

\author[Midooka T. et al.]{
Takuya Midooka$^{1,2}$,\thanks{E-mail: midooka@ac.jaxa.jp}
Yasuharu Sugawara$^{1}$, 
Ken Ebisawa$^{1,2}$
\\
$^{1}$Institute of Space and Astronautical Science (ISAS), Japan Aerospace Exploration Agency (JAXA), 3-1-1 Yoshinodai, Chuo-ku, \\Sagamihara, Kanagawa 252-5210, Japan\\
$^{2}$Department of Astronomy, Graduate School of Science, The University of Tokyo, 7-3-1 Hongo, Bunkyo-ku, Tokyo 113-0033, Japan \\
}
\date{Accepted XXX. Received YYY; in original form ZZZ}

\pubyear{2018}

\begin{document}
\label{firstpage}
\pagerange{\pageref{firstpage}--\pageref{lastpage}}
\maketitle

\begin{abstract}
WR~125 is considered as a {\it Colliding Wind Wolf-rayet Binary} (CWWB), from which the most recent infrared flux increase was reported between 1990 and 1993.
We observed the object four times from November 2016 to May 2017 with {\it Swift} and {\it XMM-Newton}, and carried out a precise X-ray spectral study for the first time.
There were hardly any changes of the fluxes and spectral shapes for half a year, and the absorption-corrected luminosity was 3.0~$\times~$10$^{33}$~erg~s$^{-1}$ in the 0.5--10.0~keV range at a distance of 4.1~kpc.
The hydrogen column density was higher than that expected from the interstellar absorption, thus the X-ray spectra were probably absorbed by the WR wind.
The energy spectrum was successfully modeled by a collisional equilibrium plasma emission, where both the plasma and the absorbing wind have unusual elemental abundances particular to the WR stars. 
In 1981, the {\it Einstein} satellite clearly detected X-rays from WR~125, whereas the {\it ROSAT} satellite hardly detected X-rays in 1991, when the binary was probably around the periastron passage.
We discuss possible causes for the unexpectedly low soft X-ray flux near the periastron. 
\end{abstract}




\section{Introduction}
Most of the Wolf-Rayet (WR) stars, massive stars with significant mass-loss, are known to be binaries \citep{Rosslowe15}.
In particular, those WR binaries which produce hot plasma from their stellar wind collision are called {\it Colliding Wind Wolf-rayet Binaries} (CWWBs). The shocked plasma has a temperature of 10$^7$--10$^8$~K and high absorption columns of 10$^{22}$--10$^{23}$ ~H~cm$^{-2}$ \citep{Schild04}. The X-ray luminosity is highly dependent on binary separations, mass-loss rates, and wind velocities \citep{Stevens92, Usov92}.
The X-ray energy spectra significantly vary with the binary orbital phase, which enable us to study orbital dependence of the plasma parameters and amount of the circumstellar absorption through spectral analysis.
In this manner, we are able to constrain the wind acceleration and the mass-loss rate from the WR star.
We have already applied this methodology to the CWWBs WR~140 \citep{Sugawara15} and WR~19 \citep{Sugawara17}, which are relatively bright with known orbital parameters. We measured variations of the circumstellar absorptions on the orbital phases, and successfully constrained the mass-loss rates from these WR stars \citep{Sugawara15,Sugawara17}.

WR~125 is considered as a CWWB, consisting of a WC7 type WR star and an O9 I\hspace{-.1em}I\hspace{-.1em}I companion star \citep{Will94}.
The orbital period is unknown, while it is reported that the infrared flux started to increase in 1990 July and reached the maximum during 1992 and 1993 \citep{Will94}.
In general, infrared brightening in the long-period CWWBs is thought to be caused by dust formation near the periastron passage in their eccentric binary orbits \citep{Williams87,Williams12}.

In 1981, the X-ray observatory {\it Einstein} detected X-rays from WR~125 for the first time \citep{Pollock87}.
The absorption-corrected luminosity was $(1.4\pm 0.4)~\times~10^{33}~{\rm erg~s^{-1}}$ in the 0.2--4.0~keV band at an assumed distance of 1.9~kpc.
As discussed in \cite{Pollock81}, the log-likelihood detection statistic $\lambda$
gives a scale of the significant detection.
In the case of {\it Einstein} IPC, $\lambda$ being greater than 3 is considered to be a significant detection. Since \cite{Pollock87} showed that $\lambda$ of WR~125 was 39.1, the detection was significant.
Later, \cite{Pollock95} claimed a marginal detection with {\it ROSAT} in 1991, where $\lambda$ was 5.8; this detection was not significant because {\it ROSAT} usually takes $\lambda$ >10 as the detection threshold.

In this paper, we present new {\it Swift} and {\it XMM-Newton} monitoring observations of WR~125, and investigate the long-term X-ray variation. In section 2 we introduce the observation and data reduction, and in section 3 we present data analysis and results. 
We discuss long-term X-ray variation of WR~125 using all the available X-ray observational results in section 4.

\begin{table*}
	\centering
	\caption{Observation logs and the count rates with {\it Swift} and {\it XMM-Newton}}
	\label{tab:obs_table}
	\begin{threeparttable}
	\scalebox{0.9}{
	\begin{tabular}{llllccc}
		\hline
		Satellite/Detector & Obs. mode & Obs. ID & Start time [UT] & Exposure time (ks) & \begin{tabular}{c}$\dot{C}$@0.3--1.5keV\tnote{a} \\(10$^{-2}$~counts~s$^{-1}$)\end{tabular}&\begin{tabular}{c}$\dot{C}$ @1.5--10.0keV\tnote{a} \\(10$^{-2}$~counts~s$^{-1}$)\end{tabular}\\
		\hline
		{\it Swift}/XRT & Photon-Counting & 00034826001& 2016-11-28T01:50 & 4.8&0.9$\pm$0.25&1.5$\pm$0.3\\
		{\it Swift}/XRT & Photon-Counting & 00034826002& 2016-12-17T13:27 & 4.7&0.6$\pm$0.20&1.7$\pm$0.3\\
		{\it Swift}/XRT & Photon-Counting & 00034826003& 2017-03-16T06:19 & 2.3&0.6$\pm$0.31&1.2$\pm$0.4\\
		{\it XMM}/EPIC & Full frame & 0794581101& 2017-05-11T09:06\tnote{b} & 21.5&5.2$\pm$0.5&15.5$\pm$0.8\\
		\hline
	\end{tabular}
	}
	\begin{tablenotes}
	\item[a] Observed count rates of each energy band. The rate of {\it XMM-Newton} is that by EPIC-pn detector. 
	\item[b] Starting time of EPIC-pn observation.
	\end{tablenotes}
	\end{threeparttable}
\end{table*}

\section{Observations and Data Reduction}
Table 1 gives the observation log and the observed count rates.
We proposed a Target of Opportunity (ToO) observation of WR~125 with {\it Neil Gehrels Swift Observatory} \citep{Gehrels04}, and three pointings were made from 2016 November 28 to 2017 March 16 for a total exposure of about 12~ksec.
The X-ray Telescope (XRT; \citealp{Burrows05}) was operated in the Photon-Counting mode.
We processed the XRT data through the {\it Swift}-XRT data product generator \footnote[1]{{\it http://www.swift.ac.uk/user\_objects/}} \citep{Evans07,Evans09}.
We produced the XRT light curves, images and spectra by using the {\it Swift}-XRT data product generator \citep{Evans07,Evans09}.

Having confirmed significant detection by {\it Swift}, we proposed a more detailed observation with {\it XMM-Newton} \citep{Jansen01}, and the observation was carried out on 2017 May 11.
The European Photon Imaging Camera (EPIC) is sensitive in the 0.2 to 12.0~keV energy range \citep{Turner01, Struder01}.
The data were reduced with SAS version 15.0.0 to obtain the filtered event files for EPIC-MOS1, 2 and pn in 0.3--10.0~keV.

Good time intervals were selected by removing the intervals dominated by flaring particle background when the single event (PATTERN $=$ 0) count rate in the $>$10~keV band was larger than 0.35~counts~s$^{-1}$ and that in the 10--12~keV band larger than 0.4~counts~s$^{-1}$ for EPIC-MOS and EPIC-pn data, respectively.
We used a circular regions of 22\arcsec~radius from the same CCD for extracting source and background events\footnote[2]{{\it https://www.cosmos.esa.int/web/xmm-newton/sas-thread-pn-spectrum}}. 
Following the SAS Data Analysis Threads\footnote[3]
{{\it https://www.cosmos.esa.int/web/xmm-newton/sas-thread-timing\\
https://www.cosmos.esa.int/web/xmm-newton/sas-thread-epic-filterbackground\\
https://www.cosmos.esa.int/web/xmm-newton/sas-thread-mos-spectrum\\
https://www.cosmos.esa.int/web/xmm-newton/sas-thread-pn-spectrum
}}, we obtained light curves and spectra.
In the following analysis, we used HEASOFT version 6.22.1 and XSPEC version 12.9.1p.

\begin{figure*}
\includegraphics[width=1.3\columnwidth, angle=270]{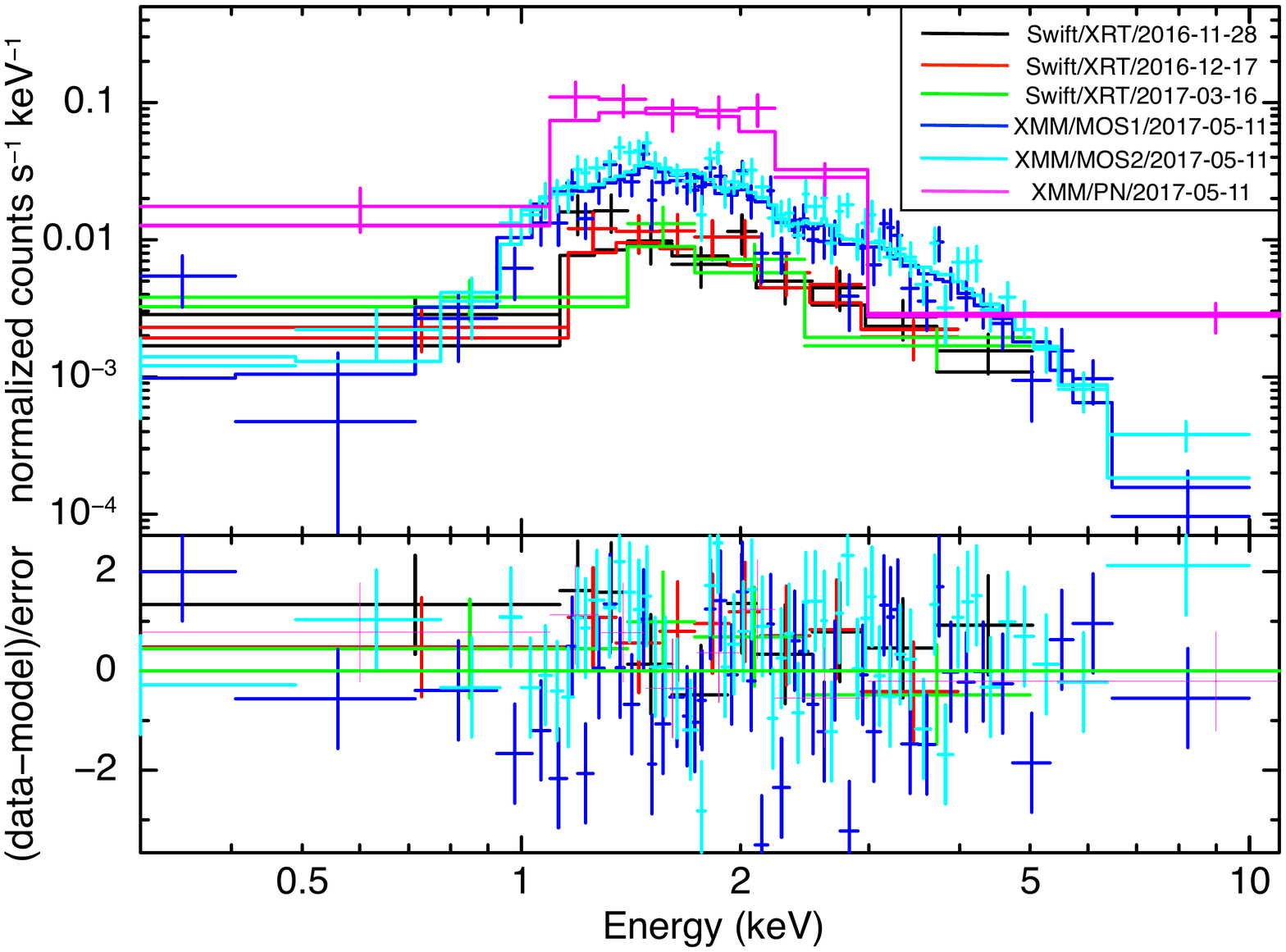}
    \caption{Spectra of WR~125 and the best-fitting models. The six spectra are fitted simultaneously. In the upper-panel, the solid lines show the best fitting models, which is {\tt TBabs*varabs*vvapec}.}
    \label{fig:fit_figure}
\end{figure*}

\begin{table}
	\centering

	\caption{The best-fitting parameter of spectral fitting }
	\label{tab:par_table}
	\begin{threeparttable}
	\scalebox{0.9}{
	\begin{tabular}{lcc}
		\hline
		\ \ \ \ \ \ \ \ \ \ \ \ Parameter &{\tt TBabs*apec}&{\tt TBabs*varabs*vvapec}\\
		\hline
		Interstellar absorption & & \\
		\ \ \ \ \ \ $N_{\rm H}~(10^{22}~{\rm cm^{-2}})$ & 1.59$\pm 0.10$ & 0.94(fixed)\\ 
		Circumstellar absorption & &\\
		\ \ \ \ \ \ $N_{\rm He}~(10^{22}~{\rm cm^{-2}})$&------- &0.16$\pm 0.06$\\
		Thin thermal plasma & &\\
		\ \ \ \ \ \ $kT$~(keV) \tnote{a}&2.33$\pm 0.16$ &2.1$^{+0.3}_{-0.2}$\\
		\ \ \ \ \ \ (Fe/He)/(Fe/He)$_\odot$&------- &0.29$\pm$0.33\\
		\ \ \ \ \ \ $E.M.~(10^{56}~{\rm cm^{-3}})$\tnote{b, c}&2.7$^{+2.1}_{-1.1}$ &1.3$^{+1.0}_{-0.5}$\\
		\ \ \ \ \ \ $F_{\rm x}~(10^{-13}~{\rm erg~cm^{-2} s^{-1}})$ \tnote{d}&7.9$^{+0.2}_{-0.3}$ &7.3$\pm 0.2$\\
		\ \ \ \ \ \ $L_{\rm x}~(10^{33}~{\rm erg~s^{-1}})$\tnote{c, e} &3.3$^{+5.5}_{-2.0}$ &3.0$^{+2.6}_{-1.2}$\\
		\hline
		\multicolumn{1}{c}{$\chi^2/{\rm dof}$} &197/135 &187/134\\
		\hline
	\end{tabular}
	}
	\begin{tablenotes}\footnotesize
	\item[a] Plasma temperature.
	\item[b] The emission measure in ${\rm cm^{-3}}$. 
	\item[c] Assuming a distance of 4.1$^{+1.2}_{-0.8}$~kpc \citep{Gaia18}.
	\item[d] The absorbed flux (0.5--10.0~keV).
	\item[e] The luminosity which is dereddened by the interstellar and \\circumstellar absorptions (0.5--10.0~keV). \\The error of the luminosity is computed using the \\maximum/minimum unabsorbed flux and distance.
	\end{tablenotes}
	\end{threeparttable}
\end{table}

\section{Data Analysis \& Results}
In the {\it Swift} of ToO observations, we detected a source at (19h 28m 15.6s, +19$^{\circ}$ 33$\arcmin$ 20.9$\arcsec$) with a 90\% radial error of 2.7$\arcsec$.
The most precise coordinate of WR~125 is (19h 28m 15.61s, +19$^{\circ}$ 33$\arcmin$ 21.53$\arcsec$) by \cite{Gaia18}, thus the detected object is certainly WR~125.
Count rates of three {\it Swift} observations were almost the same (Table 1).

We used XSPEC to analyze X-ray spectra. We made three energy spectra corresponding to three {\it Swift} observations. For {\it XMM-Newton}, we made MOS1, MOS2 and pn energy spectra, separately.
We grouped three {\it Swift} spectra every 10 counts per bin and three {\it XMM-Newton} spectra (MOS1, MOS2, pn) every 15 counts per bin.

We set the solar abundance by \cite{wilms00}, and fitted the spectra using a simple model ({\tt TBabs*apec}), where an emission spectrum from collisionally-ionized diffuse gas is affected by the interstellar absorption \citep{Smith01}. First, we fitted the six spectra separately, and found that there were hardly spectral variations. 
Consequently, we fitted the six spectra simultaneously.
The left-hand side of Table 2 shows the best-fit parameters using the simple model ({\tt TBabs*apec}).

Now we estimate $N_{\rm H}$ of the interstellar absorption from optical extinction.
According to a catalogue of Galactic WR stars \citep{Van01}, $A_v$ is 6.68 mag for WR~125. Consequently, $N_{\rm H}$ was estimated as $0.94~\times~10^{22}~{\rm cm^{-2}}$ using the following equation \citep{Vuong03},
\begin{eqnarray*}
N_{\rm H}=1.41\times A_v\times 10^{21}~{\rm cm^{-2}}.
\end{eqnarray*}
Meanwhile, our best-fit column density was $1.59~\times~10^{22}~{\rm cm^{-2}}$.
Therefore, we suppose that X-rays were further absorbed by the WR wind. 

Next, we introduced another absorption model ({\tt varabs}), in which elemental abundance is variable, in order to take the additional circumstellar absorption into account, fixing $N_{\rm H}$ of {\tt TBabs} at the expected interstellar value ($0.94~\times~10^{22}~{\rm cm^{-2}}$). We also changed {\tt apec} to {\tt vvapec} in order to specify abundances of the collisional equilibrium plasma and set H abundance to zero.
We took the C, O and Ne abundances of WR~90, which is another WC7-type WR single star and fixed other chemical abundances to the unknown Fe abundance. 
Since hydrogen is depleted, we specified the C, O and Ne abundances relative to He, as \citep[i.e. (C/He)$_*$/(C/He)$_{\odot}$ = 101.7, (O/He)$_*$/(O/He)$_{\odot}$ = 5.98 and (Ne/He)$_*$/(Ne/He)$_{\odot}$ = 3.81; ][]{Dessart00}.
Abundances of H and N were set to zero, which is expected for WR~125 being a WC-type WR star, and the abundances of the emission ({\tt vvapec}) and absorption ({\tt varabs}) components were made equal. We fitted the six spectra simultaneously, allowing only the He abundance of {\tt varabs} and the Fe abundance and {\it kT} of {\tt vvapec} to be free parameters.

The right-hand side of Table 2 shows the best-fit parameters using the more sophisticated model ({\tt TBabs*varabs*vvapec}).
$\chi^2/{\rm dof}$ was 187/134(= 1.40), slightly better than that of the simple model.
According to \cite{gagne12}, absorption-corrected luminosities and temperatures of CWWBs range from 10$^{31}$ to 10$^{35}~{\rm erg~s^{-1}}$ and from 1 to 4 keV.
The best-fit luminosity and plasma temperature of WR~125 were found within these ranges.
The energy spectra and the best-fit models are shown in Figure 1.

\section{Discussion}
We detected persistent X-ray emission from the Colliding Wind Wolf-rayet Binary WR~125 with {\it Swift} and {\it XMM-Newton} in a series of four observations carried out in 2016-2017, following a clear detection with {\it Einstein} in 1981 and a marginal detection with {\it ROSAT} in 1991.
No significant flux/spectral changes were found throughout the first observation in 2016 November to the last one in 2017 May.
We suppose that the orbital period may be longer than 24 years, considering that the last reported periastron passage (expected from the near infrared flux increase) was in 1993 \citep{Will94}, and there was no flux increase reported since then.

We carried out X-ray spectral analysis in 0.3--10~keV from WR~125 for the first time.
From the spectra analysis, we found that the column density was probably increased by WR~125's stellar wind component, and the plasma parameters (luminosity and temperature) were not so extreme values among WC-type WR binaries \citep{gagne12}.

We carefully looked into the archival data of {\it Einstein} and {\it ROSAT}. {\it Einstein} data by Imaging Proportional Counter (IPC) instrument was sensitive in the 0.4 to 4.0~keV energy range.
It was obtained in 1981 April 9 (sequence No. 8680), and the count rate was 0.0122$(\pm 0.0028)~{\rm counts~s^{-1}}$, which is considered as a significant detection \citep{Pollock87, Harris96}.
On 1991 October 28, {\it ROSAT} data was taken by Position Sensitive Proportional Counters (PSPC) instrument in 0.1-2.0 energy range for an exposure of 2105 seconds (sequence ID RP500042N00).
As a result of scrutinizing the {\it ROSAT} data, we conclude that there was no meaningful X-ray detection from WR~125; the WR~125 count rate was less than that of the dimmest point source significantly detected in the field-of-view ($5.0~\times~10^{-3}~{\rm counts~s^{-1}}$).

With WebPIMMS (ver. 4.9)\footnote[4]{{\it https://heasarc.gsfc.nasa.gov/cgi-\\*bin/Tools/w3pimms/w3pimms.pl}}, we converted the {\it Einstein} count rate into the flux in the 0.5--10.0~keV energy range assuming $N_{\rm H}$ = 1.59~$\times~10^{22}~{\rm cm^{-2}}$, plasma temperature = 2.16~keV and 1.0 solar abundance in {\tt APEC} model based on {\it Swift} and {\it XMM-Newton} (see above).
The converted flux was 7.3$(\pm 1.7)~\times~$10$^{-13}$~{\rm erg~s}$^{-1}$~{\rm cm$^{-2}$}.
We also converted the {\it ROSAT} upper-limit count rate ($5.0~\times~10^{-3}~{\rm counts~s^{-1}}$) into the flux in the 0.5--10.0~keV range; it was 4.2~$\times$~10$^{-13}$~{\rm erg~s}$^{-1}$~{\rm cm$^{-2}$}.

\begin{table}
	\centering
	\caption{Observed flux with {\it Einstein}, {\it ROSAT}, {\it Swift} and {\it XMM-Newton}}
	\label{tab:flux_table}
	\begin{threeparttable}
	\scalebox{0.9}{
	\begin{tabular}{llc}
		\hline
		Obs. date & Satellite/Detector&Observed flux \tnote{a}\\
		(yyyy.mm)&&(10$^{-13}$~erg~s$^{-1}$~cm$^{-2}$)\\
		\hline
		1981.04&{\it Einstein}/IPC&7.3$\pm 1.7$\\
		1991.10&{\it ROSAT}/PSPC&<~4.2\\
		2016.11--2017.05&{\it Swift}/XRT\&{\it XMM}/EPIC&7.9$^{+0.2}_{-0.3}$\\ \hline
	\end{tabular}
	}
	\begin{tablenotes}\footnotesize
	\item[a] We converted each count rate to the flux in the 0.5--10.0~keV \\range using model {\tt APEC} with WebPIMMS.
	\end{tablenotes}
	\end{threeparttable}
\end{table}

Table 3 shows the observed flux with {\it Einstein}, {\it ROSAT} and {\it XMM-Newton} ({\it Swift}) in the 0.5--10.0~keV energy range, where the flux observed with {\it ROSAT} in 1991 was obviously the lowest. 
Meanwhile, the infrared flux was increasing then, thus WR~125 was probably in the periastron passage \citep{Will94}. 
Namely, contrary to the expectation that X-ray luminosity of the internal shock layer is inversely proportional to the binary separation \citep{Usov92}, the {\it ROSAT} observation suggested that the soft X-ray luminosity decreased at the periastron. 
Coincidentally, a similar soft X-ray decrease at the periastron was also observed from WR 22 \citep{Gosset09}, Eta Carinae \citep{Cocoran10}, WR~140 \citep{Sugawara15} and WR~21a \citep{Gosset16}.

In order to understand the observed low luminosity near the periastron, we examined three possibilities:
First, there may be a chance that the WR star or the companion star coincidentally fully occulted the colliding wind region in 1991 exactly when {\it ROSAT} observed the source.
Since the orbital inclination angle is not restricted at all from IR or optical observations, it is difficult to estimate the eclipse possibility.
In any case, even though the orbit of WR~125 has a high inclination, eclipse may not be expected in X-ray band.
Actually, a total X-ray eclipse was never reported up to now in any CWWBs; for example, WR~20a does not show any eclipses \citep{Naze08} and V444 Cyg shows only a partial one (e.g. \citealp{Lomax15}) despite of their high inclination angles.
Therefore, the first possibility may be low.

Second, soft X-ray from the colliding wind region may have been heavily absorbed by the WR star wind, while intrinsic X-ray luminosity of WR~125 is not significantly variable.
While {\it ROSAT}/PSPC was sensitive to X-rays only between 0.1 and 2.0~keV, {\it Swift} and {\it XMM-Newton} are respectively sensitive in the 0.3 to 10.0~keV and 0.3 to 12.0 keV energy ranges.
Therefore, it might be possible that {\it ROSAT} was not able to detect the soft X-rays if significantly absorbed by the WR star wind.
When we increased $N_{\rm H}$ from the best-fit $1.6~\times~10^{22}~{\rm cm^{-2}}$ to $1.0 \times 10^{23}~{\rm cm^{-2}}$ assuming the same intrinsic luminosity and the spectra determined by {\it Swift/XMM}, we found it impossible to detect WR~125 using {\it ROSAT}/PSPC.
However, in fact, there are few observations that CWWB has such a high column density \citep{Rauw00, Schild04, Sugawara15}.
Then, we examined requirements that WR~125 column density would reach $1.0 \times 10^{23}~{\rm cm^{-2}}$.
According to \cite{Pollock05}, column density of a spherically symmetric WR wind at a distance $R$ from the WR star surface along the line of sight can be written as
\begin{eqnarray*}
N_{\rm H}(R, \phi, i) \sim 4.31\times10^{23} \dot{M}_{-6} \mu^{-1} v_8^{-1} (R_*/R_{\odot})^{-1} \\ \times~(\gamma/{\rm sin}\gamma) \int_{R/R_*}^\infty x^{-2} (1-1/x)^{-2} dx~{\rm cm^{-2}},
\end{eqnarray*}
where cos$\gamma$ = cos$\phi$ sin$i$, mass-loss rate $\dot{M}_{-6} = 10^{-6} M_{\odot}$ yr$^{-1}$, wind velocity $v_8 = 1000$ km s$^{-1}$, $\phi$ and $i$ are the orbital azimuthal and inclination angle, and $R_*$ is the WR stellar radius.
In the case of WR~125, we used typical physical conditions in the WC type WR wind $\dot{M} = 2~\times~10^{-5}~M_{\odot}~$yr$^{-1}$, $v_{\infty}$ = 2000 {km s$^{-1}$} \citep{Dessart00}, $R_* = 6.0R_{\odot}$ \citep{Koesterke95}, and mean atomic weight for nucleons $\mu$ = 6, which was estimated by the He, C, O and Ne abundances.
Since we cannot constrain other orbital parameters, we assumed $\phi = 180^{\circ}$, which gives the maximum column density.
We examined two cases for different location of the X-ray emitting plasma, (1) R = 0.5AU and (2) R = 1.0AU.
As a result, we found that only when inclination $i$ is more than 78$^{\circ}$ in situation (1) or $i$ is more than 84$^{\circ}$ in situation (2), the column density reaches $1.0 \times 10^{23}~{\rm cm^{-2}}$.
Therefore, it is possible to attain $N_{\rm H} = 1.0 \times 10^{23}~{\rm cm^{-2}}$ only under the very limited circumstances with particular binary separation, orbital inclination angle and azimuthal angle. 

Third, size of the X-ray emitting (colliding wind) might be reduced near the periastron under some circumstantial conditions.
For example, one possibility is lack of the enough acceleration in the O-star wind.
In general, wind momentum of the WR star overwhelms that of the O-star, so that the colliding region almost reaches the O-star surface. 
Consequently, near the periastron, O-star wind may not have sufficient space to reach its terminal velocity before entering the shock region, and collides with the WR star wind before reaching the terminal velocity; this will lead to reduction in the wind momentum fluxes (e.g. \citealp{Luo90}; \citealp{Stevens92}; \citealp{Myasnikov93}).
Other possibilities are radiative inhibition and radiative braking, which can be obstacles of wind-acceleration (e.g. \citealp{Stevens94}, \citealp{Gayley97}).
The radiative inhibition is a process where the acceleration of each wind is reduced by the radiation from its companion star.
The radiative braking describes a scenario in which the WR wind is slowed after reaching a large velocity. 
These mechanisms require small binary separations; for example, the smallest separation in V444 Cyg is 35.97 R$_{\odot}$ (0.33AU) \citep{Eris11}.
If binary separation of WR~125 is sufficiently small, these processes can slow down the wind velocity significantly, and reduce size of the colliding wind region, decreasing the X-ray flux.
With a hydrodynamical simulation, it is suggested that X-ray flux could even disappear due to a full disruption of the colliding wind region \citep[e.g.][]{Parkin11}.
In conclusion, we suppose that the significant low soft X-ray flux in 1991 was likely to be a consequence of mixture of the second and third possibilities.

In summary, we have confirmed a long-term X-ray variation from WR~125 over 36 years for the first time using four X-ray satellites. Still, WR~125 has many unknown aspects, even its orbital period.
If we can determine the orbital parameters precisely in future, we may understand reason of the significantly low luminosity in 1991.  
According to \cite{Will92}, extinction of the non-thermal radio emission is expected to increase by the dense WR wind material just before the dust formation.
Thus, we suppose that significant change of the radio flux may become a sign of the periastron passage.
We propose multi-band monitoring observations of WR~125 including radio, in order to determine the orbital parameters and clarify the wind parameters.

\section*{Acknowledgements}
We are grateful to the anonymous referee for the comprehensive review and many constructive comments.
This research has made use of data and software provided by the High Energy Astrophysics Science Archive Research Center (HEASARC), which is a service of the Astrophysics Science Division at NASA/GSFC and the High Energy Astrophysics Division of the Smithsonian Astrophysical Observatory.
We acknowledge the use of public data from the {\it Swift} data archive and the UK Swift Science Data Center at the University of Leicester. 
This study was based on observations obtained with {\it XMM-Newton}, an ESA science mission with instruments and contributions directly funded by ESA Member States and NASA.
This research was partially supported by JSPS KAKENHI Grant Numbers JP16K17667 (Y.S.), JP16K05309 (K.E.).




\bibliography{wr125}



\bsp	
\label{lastpage}
\end{document}